\documentclass[conference]{IEEEtran}
\usepackage{threeparttable}
 
\usepackage{multicol}
\usepackage{color,soul}
\usepackage{threeparttable}

\usepackage{cite}

\ifCLASSINFOpdf
   \usepackage[pdftex]{graphicx}
   \DeclareGraphicsExtensions{.pdf,.jpeg,.png}
\else
\fi
\usepackage{amsmath}
\usepackage{url}

\usepackage[]{fancyhdr} %
\newcommand{\changefont}{\fontsize{9}{9}\selectfont}
\ifCLASSOPTIONcompsoc
  \usepackage[caption=false,font=normalsize,labelfont=sf,textfont=sf]{subfig}
\else
  \usepackage[caption=false,font=footnotesize]{subfig}
\fi
\usepackage{url}

\usepackage{float}
\usepackage{xcolor}
\usepackage{color,soul}
\usepackage{multirow}

\IEEEoverridecommandlockouts
\begin{document}

%
\title{Identification of Substation Configurations in Modern Power Systems using Artificial Intelligence}

\author{\IEEEauthorblockN{Dulip~Madurasinghe, \emph{Student Member, IEEE} \\ Real-Time Power and Intelligent Systems Laboratory}
\IEEEauthorblockA{Holcombe Department of Electrical \\ and Computer Engineering\\Clemson University\\
Clemson, SC 29634, USA\\
dtmadurasinghe@ieee.org}
\and
\IEEEauthorblockN{Ganesh~K.~Venayagamoorthy, \emph{Fellow, IEEE} \\Real-Time Power and Intelligent Systems Laboratory}
\IEEEauthorblockA{Holcombe Department of Electrical \\ and Computer Engineering\\Clemson University\\
Clemson, SC 29634, USA\\
gkumar@ieee.org}

\thanks{This work is supported in part by the US National Science Foundation (NSF) under grants 2131070 and the Duke Energy Distinguished Professor Endowment Fund. 

Any opinions, findings and conclusions or recommendations expressed in this material are those of the author(s) and do not necessarily reflect the views of National Science Foundation and Duke Energy.}}


%





\maketitle
\thispagestyle{fancy}
\pagestyle{fancy}


\begin{abstract}

Power system transmission network topology is utilized in energy management system applications. Substation configurations are fundamental to transmission network topology processing. Modern power systems consisting of renewable energy sources require reliable and fast network topology processing due to the variable nature of wind and solar power plants. Currently used transmission network topology processing, which is based on the relay signals communicated through SCADA is not highly reliable or highly accurate. Substation configuration identification (SCI) for different substation arrangements including main and transfer bus arrangement (MTBA), ring bus arrangement (RBA), and single bus arrangement (SBA) is investigated. Synchrophasor measurement based SCI for functional arrangements (FA) using artificial intelligence (AI) approaches is proposed in this paper. This method improves monitoring FA. Typical results for MTBA, RBA and SBA substation configuration identification is presented. A modified two-area four-machine power system model with two grid connected solar PV plants consisting of MTBA, RBA and SBA is simulated on real-time digital simulator. AI based SCI is shown to accurately identify all possible FAs for the three substation arrangements under any operating condition.

\end{abstract}


%
\IEEEpeerreviewmaketitle

\section{Introduction}\label{sec:intro}

Power system is a large, geographically distributed single circuit that supply power to the consumers from the generation stations. Modern power system is highly dynamic due to the nature of renewable energy sources (RES), distributed energy resources, etc. Thus, modern power system control requires high resolution real-time system monitoring and processing capability. The control center is responsible for the reliable and secure power system operation. Thus, control center requires the knowledge of each and every component's connectivity to the transmission network. Power system transmission network topology processing (TNTP) is used in the control center to understand the power system transmission network component connectivity. TNTP defines the node-breaker model (NBM) based on status of breakers located in substations~\cite{powerworld_1}. The NBM is converted to a dynamic bus-branch model (BBM) to be used in the energy management system (EMS) applications. Substation Bus sections or branches with very low or zero impedance are eliminated under this conversion to facilitate efficient numerical calculations~\cite{powerworld_2}.

BBM derived by the TNTP is used by EMS online applications including state estimation~\cite{powerworld_2},~\cite{18},~\cite{20}, power flow~\cite{19}. At the same time, reliable BBM is important for energy market application such as locational marginal pricing~\cite{8}, security constrained optimal dispatch and optimized renewable integration, unit commitment~\cite{26} . Power system planning division uses TNTP to conduct contingency analysis and stability analysis~\cite{powerworld_2}, where online BBM can be used to decrease the gap between planning simulation results and the practical results~\cite{powerworld_2}.

Currently, TNTP uses the relay signals collected by the remote terminal units in the substation about breaker status, which communicated to the control center via SCADA to derive the power system TNTP~\cite{powerworld_2}. Thus, the substation configuration is the fundamental for TNTP. The existing TNTP is not precise~\cite{9}. At the same time, breaker malfunction or false data injection by a man-in-the-middle attack~\cite{4} can easily compromised the derived TNTP in the control center. Which cause erroneous in results obtained by the applications based on TNTP. There has been numerous research focus on this problem. As described in~\cite{22}, graph methods can be integrated to derive the changes in the TNTP. As proposed in~\cite{23}, novel wireless sensor network can be establish to identify the branch status, which can be used to derive the TNTP. Thus, an improved TNTP or an validation for the existing TNTP is required. This can be done either communicating the breaker status through a secure communication channel or data-driven TNTP based on the measurements communicated through a secure communication channel respectively. Although implemented PMUs has the capability of integrating the digital relay signals of breakers status to it's data frame~\cite{21}, implementing such infrastructure in the existing synchrophasor network will be expensive. Thus, a redundant TNTP that use the measurements communicated through secure communication channel with only addition of computational nodes to the system without modification to the infrastructure is ideal.

To enhance the situational awareness of the power system, Phasor Measurement Units (PMUs) are introduced~\cite{3} in the substations to enable high resolution real-time measurements. PMUs measure phasor quantities at a higher rate. Typically, 30Hz reporting frequency is used. This data set is a representation of the power system state, which open possibilities to investigate the system online and implement TNTP based solely on the measurements. Although there are proposed methods to improve the TNTP by introducing measurement based TNTP~\cite{10}-~\cite{11} most of this are lacking in-depth investigation of substation configurations. The substation configuration identification is studied in~\cite{16},~\cite{24}, which are still rely on the breaker signals, which is again depends on the existing method of TNTP. This study is an effort to investigate the different substation configurations. Substation configuration identification (SCI) is the first level of TNTP. 

\begin{figure*}[!ht]
    \centering
    \captionsetup{justification=centering,margin=2cm}
    \includegraphics[width=\textwidth]{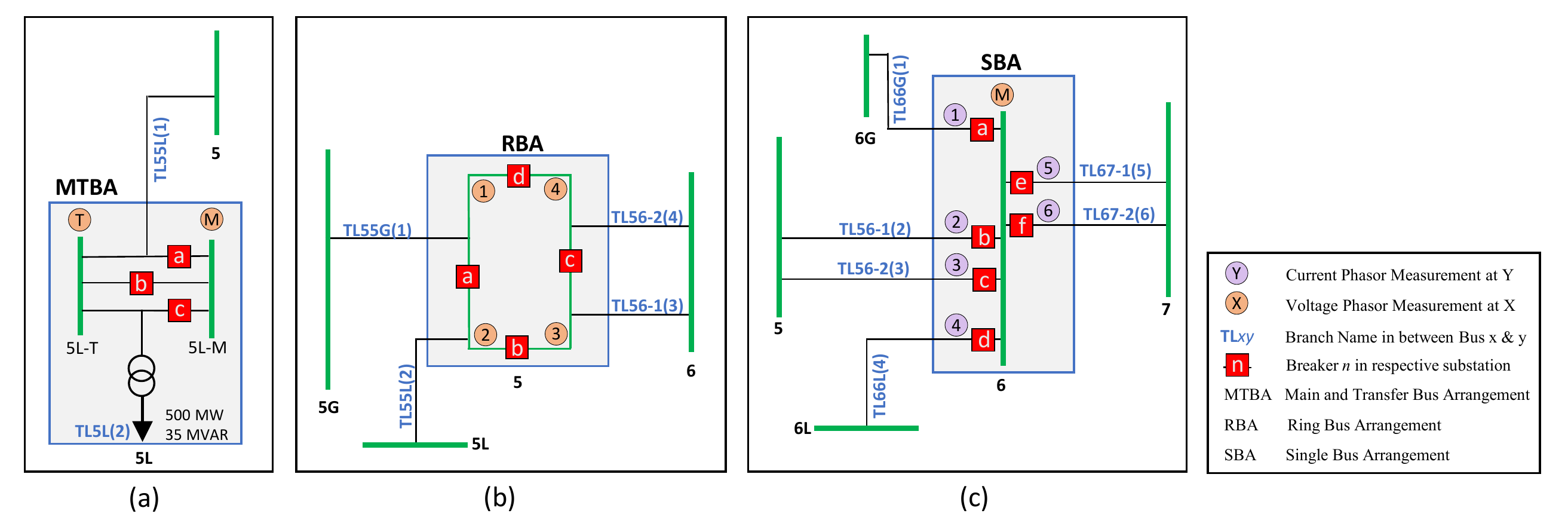}
    \caption{Substation Circuit Arrangements: a) Main and Transfer Bus Arrangement (MTBA), b) Ring Bus Arrangement (RBA) and c) Single Bus Arrangement (SBA).}
    \label{fig:substations}
\end{figure*}

High voltage substations are established in critical locations of the power system transmission network to enable control of power flow and monitoring of system states to supply power securely. Connectivity of power system critical components such as generators, transmission lines, transformers, loads, etc. can be controlled in the substations by operating the breakers accordingly. Power system measurement instrumentation such as PMUs are deployed in the substations to monitor branch current, bus-section voltage, power flow and frequency. There are several substation arrangements~\cite{mcdonald}. In this study, three such substation arrangements are considered. They are main and transfer bus arrangement (MTBA), ring bus arrangement(RBA) and single bus arrangement (SBA) as shown in Fig.~\ref{fig:substations}. Selection of the particular substation arrangement to extend in the power system transmission network is based on the type of connections, level of required reliability and the expenditure capability. Furthermore explanation on selected three substation arrangement types are explained in section II.

The demand side power requirement is becoming complex. The utility grid requires modernization to be compatible with complex demand requirement. TNTP is one application that will come under this modernization~\cite{14}. The future power grid will have higher number of switching operations and frequent topology changes. This is driven by the rapid integration of renewable source based generation which is intermittent and the accumulation of commercial micro-grids, which islanding will be a standard frequent operation. Artificial intelligence (AI) algorithms can be used to establish a real-time TNTP~\cite{11}. Furthermore, AI techniques can be used to provide enhanced situational awareness with PMU data~\cite{25},~\cite{12}. In this study, AI based SCI for MTBA, RBA and SBA is proposed. Use of data-driven AI based approach is new to SCI. The authors considered this approach due to following considerations,

\begin{itemize}
    \item Future power system transmission network is expected to handle more frequent topology changes. Thus, more intelligent AI based approach is preferred.
    \item The proposed AI based approach is scalable to handle substation expansions with growing power system complexity.
    \item Operation and management of modern power systems will be more dependent on real-time complex data and thus, AI based approaches are more robust for substation configuration identification.
\end{itemize}

The SCI is based on either knowledge based logical decision making (LDM) or neural network (NN). An LDM or NN  based SCI for each substation type is selected based on complexity of the arrangement and efficiency of the algorithm.

SCI is tested using PMU data from a real-time simulation of MTBA, RBA and SBA. Section II, elaborates on the substation configurations. Section III, explains the AI based SCI methodology. In section IV, results of all possible substation configurations for the MTBA, SBA and RBA is presented. Section V concludes the study by elaborating on the findings and future directions of the study.


\section{Substation Configuration}

A substation consists of branches, bus-sections, protection equipment (relays), measurement instruments (PMUs) and switching equipment (Breakers, Isolators). Each of these equipment is required in the power system control. TNTP is defined based on the status of each breaker established in all substations. The breaker status is identified using the relays and the relay signals are communicated to the control center, which is eventually used to established the TNTP. The substation configurations can be analyzed as follows, 

\smallskip
\begin{enumerate}
  \item Component Arrangement (CA).
  \item Functional Arrangement (FA).
\end{enumerate}
\smallskip

CA explains the physical connectivity of each branch and bus-section of the substation. CA is the NBM in the substation context. Each breaker status is used to define the CA of the substation. FA is explained by the fundamental electrical circuit nodal analysis theorems. FA is the BBM in the substation context. The FAs are a subset of the CAs. 

Since the power system TNTP is highly dependent on the reliable relay signal communication, an fail-safe redundancy or a validation is recommended. Use of synchrophasor network measurements to establish an intelligent data-driven BBM is ideal to overcome this issue. A novel BBM can be defined by integrating all substation FAs. This novel BBM can be used as a replacement or to validate the current BBM, which is derived using TNTP based on relay signals communicated to the control center~\cite{powerworld_2}. Thus, extension of the proposed FA for SCI in this study can be used to improve the reliability of the existing BBM of the network and the EMS applications. The remaining subsections describe the MTBA, RBA and SBA arrangements and elaborate on FAs.

\subsection{Main and Transfer Bus Arrangement (MTBA)}

The MTBA substation arrangement design to connect all the branches in between main bus and transfer bus. At any point of active operation the main bus can be taken out of service and the substation operation can be continued without interruptions using only the transfer bus. Although the transfer bus is typically used in a main bus maintenance, it can be used as a redundancy for the main bus in the substation under emergency operation. PMUs are installed to measure the bus voltages. MTBA can have $2^n$ (where n = number of breakers in the substation) CAs and only two FAs. A load transfer substation of MTBA with two branches is shown in the Fig.~\ref{fig:substations} (a).

\subsection{Ring Bus Arrangement (RBA)}

\begin{figure}[h!]
\centering
\includegraphics[width=3in]{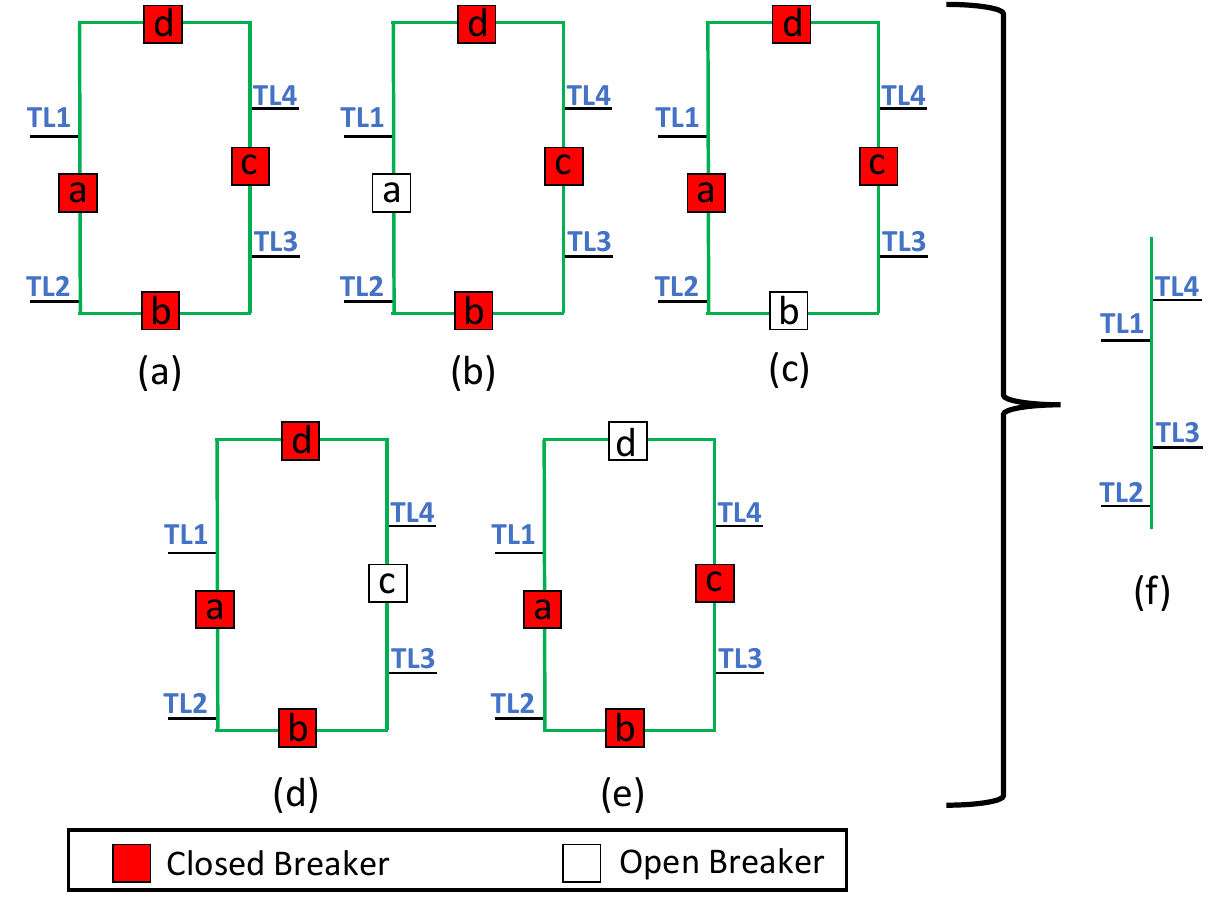}
\caption{Four-node RBA. Functional arrangement of the component arrangements in (a) - (e) is shown in (f).}
\label{fig:func_vs_struct}
\end{figure}

The RBA consists of bus-sections connected through circuit breakers. RBA is a reliable bus arrangement. Any single breaker malfunction does not affect the substation operation. Combination of multiple circuit breakers operations are used to disconnect and reconnect high voltage power system components such as generators, transmission lines, transformers, etc. to the substation and change the power routing pattern. The common practice by the utilities to measure each current through connected branches and the voltages of the bus sections at the connected node of the ring bus arranged substation. Thus, current phasor measurements of each branch connecting to the RBA substation and voltage phasor measurements of each bus section is available at the substation and the control center. RBA have $2^n$ (where n = number of bus sections or breakers in the substation) CAs and ($2^n-n-1$) FAs. The RBA test case of this study, is a four-node RBA substation with PMU measurement points is shown in the Fig.~\ref{fig:substations} (b). An example of CAs and FAs of a four-node RBA is shown in Fig.~\ref{fig:func_vs_struct}.

\subsection{Single Bus Arrangement (SBA)}

The SBA is the fundamental substation arrangement. Reliability is very low due to lack of redundancy under breaker failure or bus fault. All the branches are directly connected to a single bus through circuit breakers. PMUs are installed to measure all branch currents and the bus voltage. Thus, both branch currents and the bus voltage is available at the control center. SBA is a special case, where each FA, and corresponding CA is unique. SBA have $2^n$ (where n = number of breakers or branches in the substation) CAs and FAs. In this study, SBA substation test case has 6 branches connected as shown in Fig.~\ref{fig:substations} (c).



\section{AI Based Substation Configuration Identification} \label{sec:meth}

\begin{figure}[h!]
\centering
\includegraphics[width=2in]{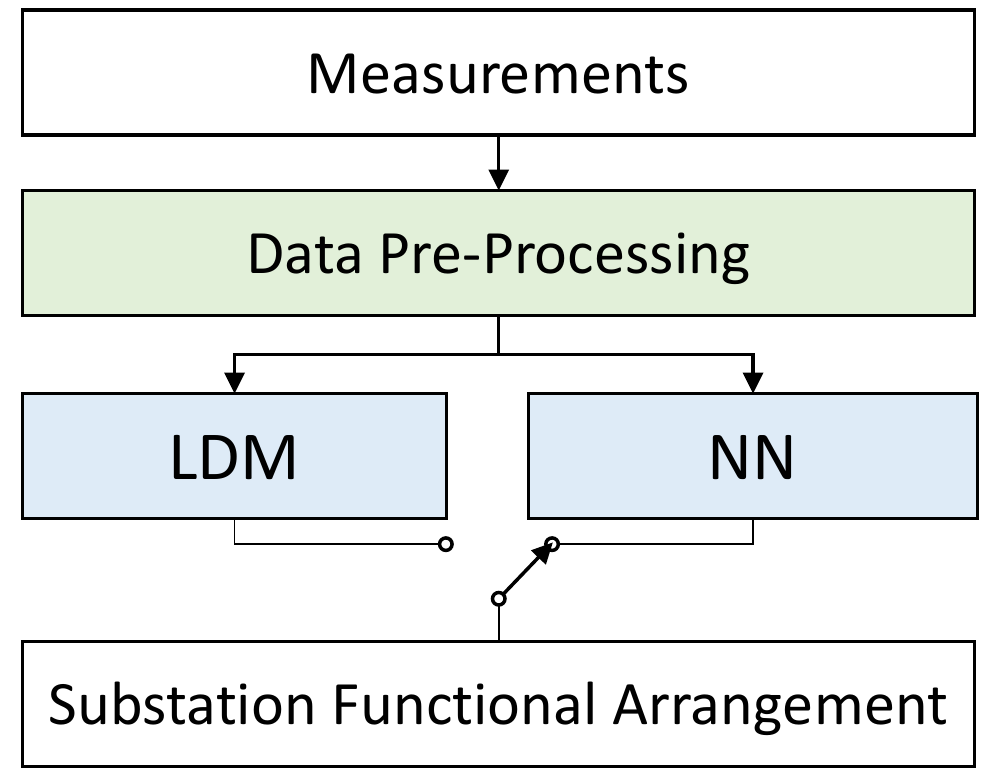}
\caption{Procedure of the proposed substation configuration identification approach.}
\label{fig:approach}
\end{figure}

In this study, two AI based algorithms are considered to identify the functional arrangements to derive substation configuration, namely, logical decision making (LDM) and neural network (NN). For each type of substation arrangement, either LDM or NN is selected considering the complexity of the problem and the arrangement based on the computation efficiency. Proposed MTBA, SBA and RBA SCIs are expected be reliable, intelligent and easy to implement. The SCI can be developed using knowledge based LDM algorithm that follows the flow diagrams shown in Figs.~\ref{fig:MTBA_flow},~\ref{fig:RBA_flow} and~\ref{fig:SBA_flow} or using NNs that are trained offline and use at any substation similarly arranged, despite of the powers system state. Training data set generation for NN approach can be done based on the simple logical algorithm that follows the highlighted section that is shown in Figs.~\ref{fig:MTBA_flow},~\ref{fig:RBA_flow} and~\ref{fig:SBA_flow}. Thus, the proposed SCI development is only requires basic substation information such as substation type, number of branches. Thus, NN based SCIs can be implemented and trained offline with a generic binary training data set and used in the field. The proposed SCI approach can be summarized as shown in Fig.~\ref{fig:approach}.

\subsection{Main and Transfer Bus Arrangement (MTBA)}

\begin{figure}[h!]
\centering
\includegraphics[width=3.49in]{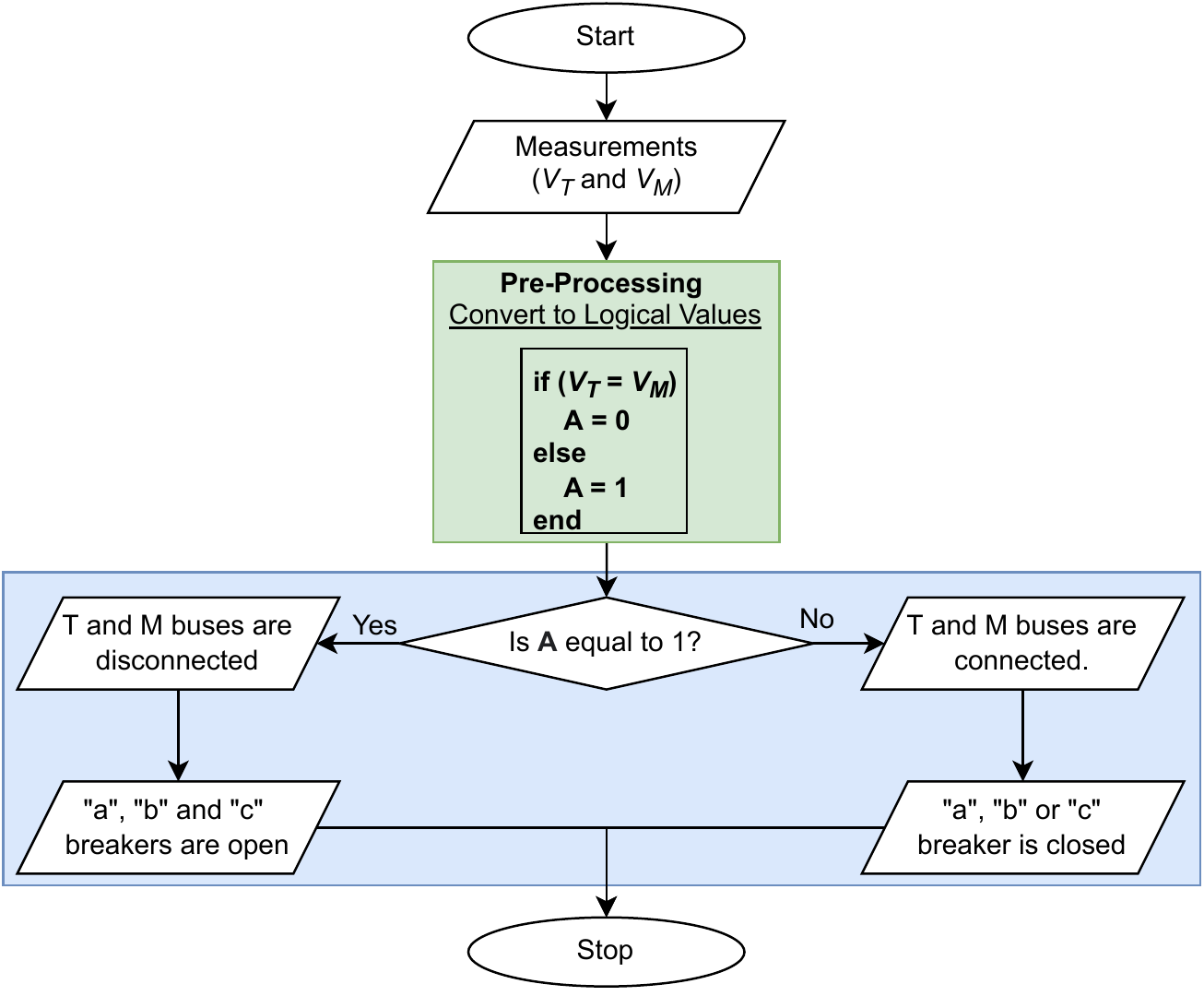}
\caption{A flow diagram for substation configuration identification of functional arrangement of MTBA type substation.}
\label{fig:MTBA_flow}
\end{figure}

\begin{figure}[h!]
\centering
\includegraphics[width=3.49in]{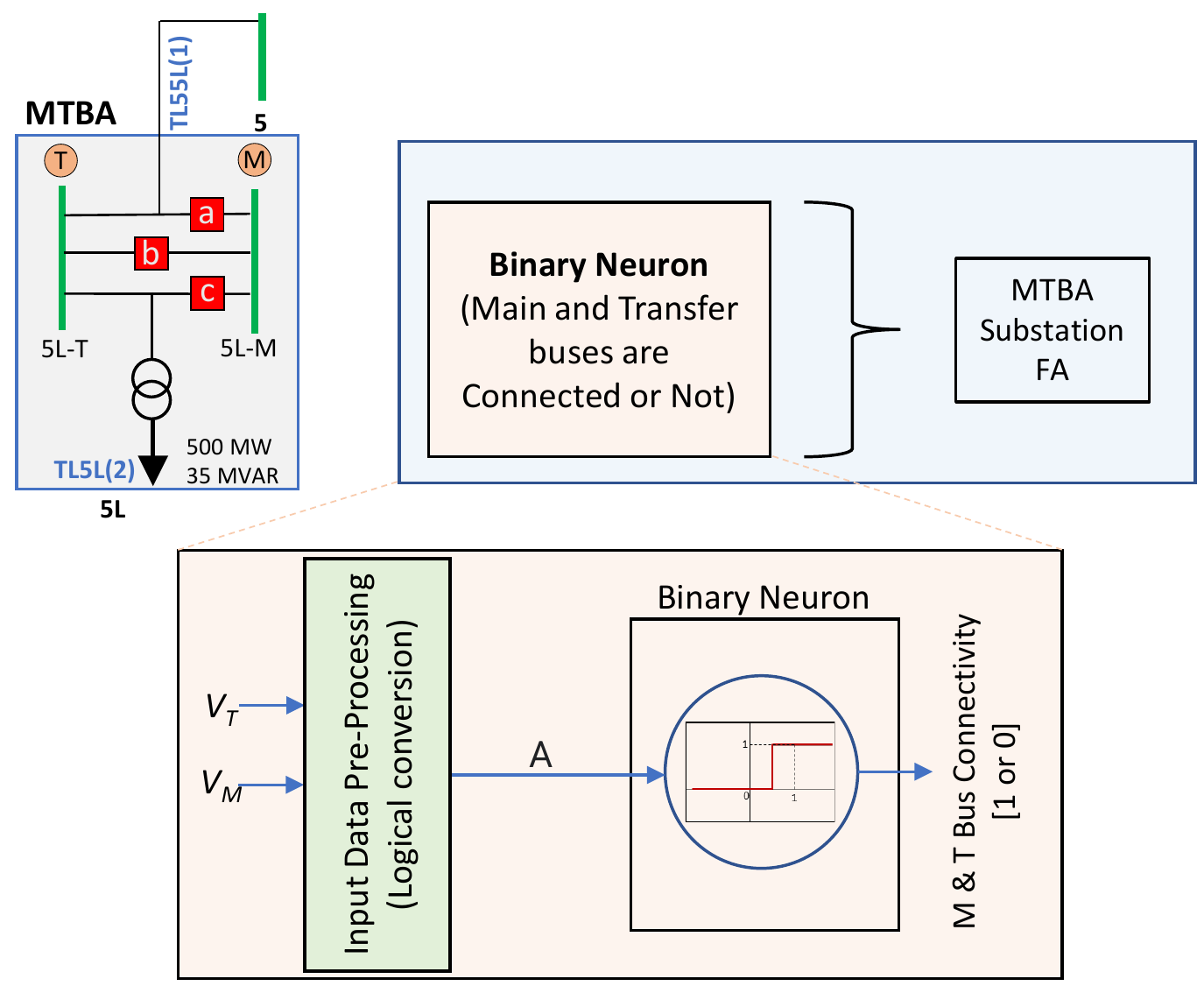}
\caption{The single binary neuron structure based substation configuration identification developed for MTBA substation shown in Fig.~\ref{fig:substations} (a).}
\label{fig:MLPNN_MTBA}
\end{figure}

MTBA substation with two branches is considered as the test case, which is shown in the Fig.~\ref{fig:substations} (a). The PMUs are in place to measure main and transfer bus voltages as shown in orange circles in the Fig.~\ref{fig:substations} (a). The two possible FAs of any MTBA substation can be inferred by using the main and transfer bus voltages. The intelligent decision procedure is shown in the Fig.~\ref{fig:MTBA_flow} flow diagram. A LDM algorithm that checks equality of main and transfer bus voltages can be developed by following the flow diagram shown Fig.~\ref{fig:MTBA_flow} or a single binary neuron can be trained using the pre-generated training data by following the flow diagram shown in Fig.~\ref{fig:MTBA_flow} as shown in Fig.~\ref{fig:MLPNN_MTBA} to identify the FA. The NN is for MTBA is a single binary neuron. This is a simple linear separable problem. Therefore a NN is not needed. Although for comparison of LDM vs NN purpose, a single neuron NN is developed.  

\subsection{Ring Bus Arrangement (RBA)}

\begin{figure}[h!]
\centering
\includegraphics[width=3.49in]{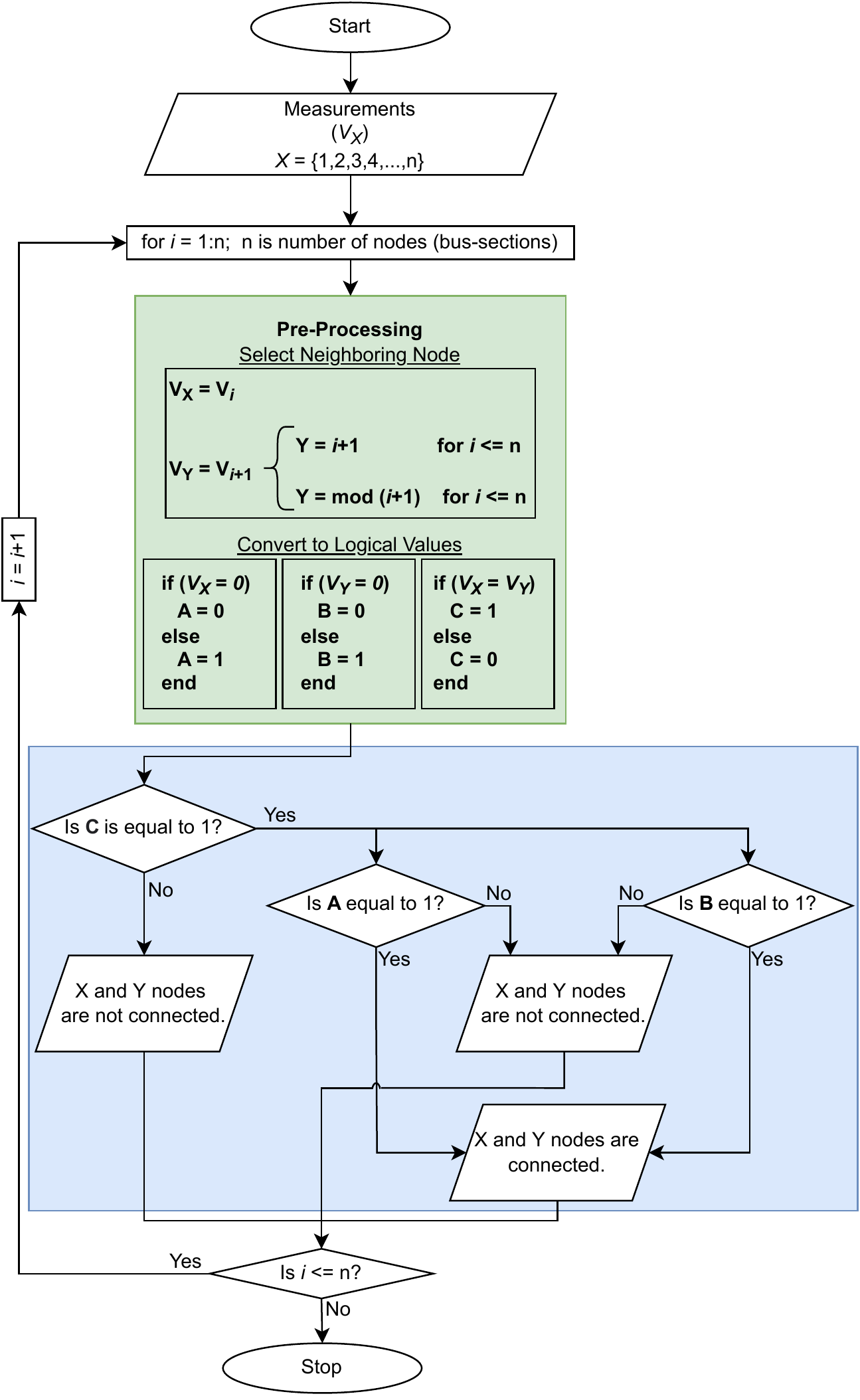}
\caption{A flow diagram for substation configuration identification of functional arrangement of RBA type substation.}
\label{fig:RBA_flow}
\end{figure}

RBA substation with four-nodes is considered as the test case, which is shown in the Fig.~\ref{fig:substations} (b). The PMUs are in place to measure each bus-section voltages as shown in orange circles in the Fig.~\ref{fig:substations} (b). Each FA of the RBA substation can be inferred based on the circuit nodal analysis using only voltage phasor measurements of the bus sections. In the substation each bus section voltage phasor is compared with neighbor bus section voltage to identify the nodal connectivity which is separated by the breaker. Bus-section voltage measurements are considered as the inputs in the RBA to identify FAs as shown in the Fig.~\ref{fig:RBA_flow} flow diagram.

\begin{figure}[h!]
\centering
\includegraphics[width=3.49in]{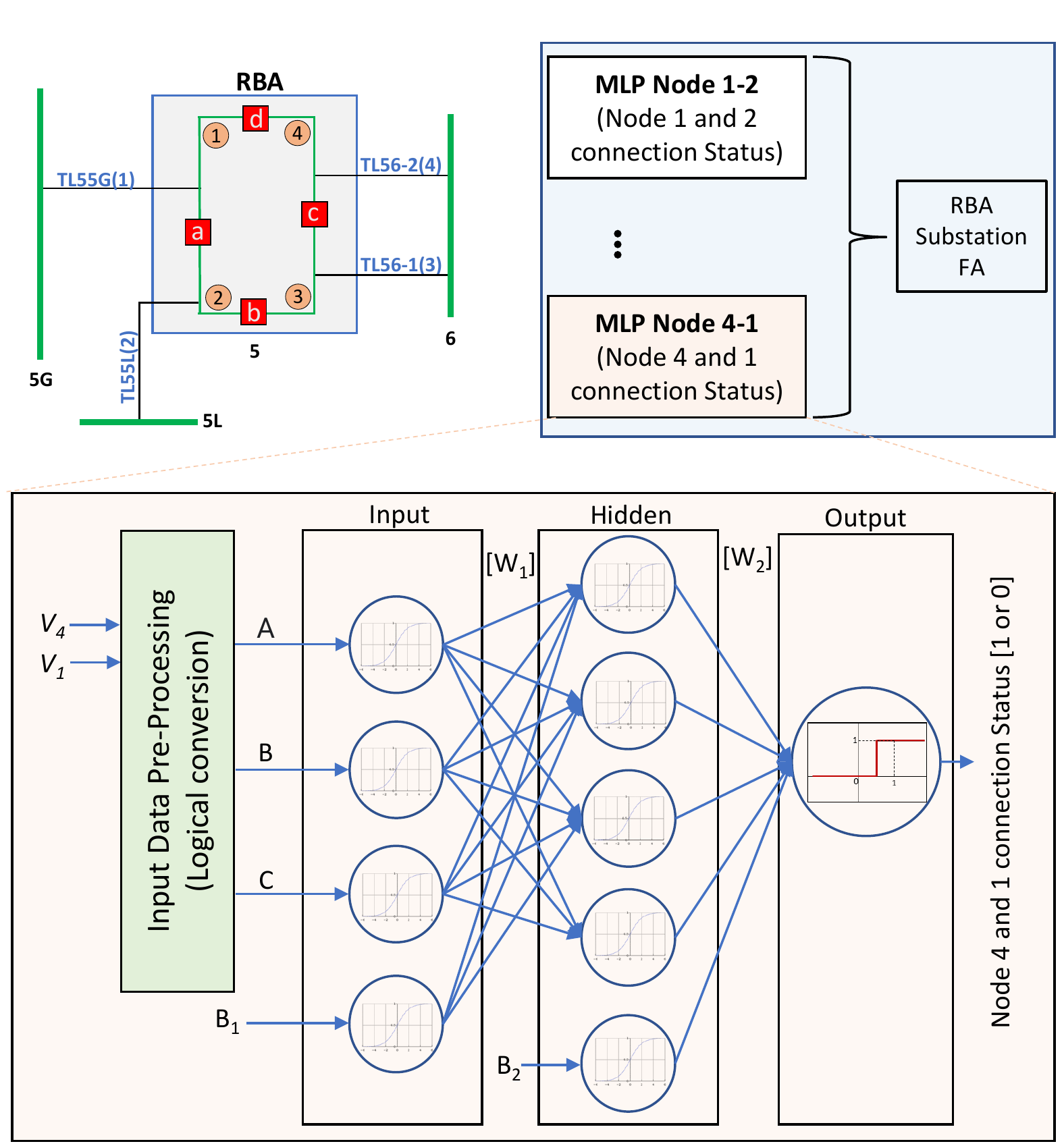}
\caption{The neural network structure of the MLP based substation configuration identification developed for a four-node RBA substation shown in Fig.~\ref{fig:substations} (b).}
\label{fig:MLPNN_RBA}
\end{figure}

X and Y are consecutive nodes in the RBA. The quantities shown are PMU measurements. For RBA substation arrangement an LDM can be setup by following the procedure shown in the Fig.~\ref{fig:RBA_flow} or a feed forward multilayer perceptron (MLP) neural network can be trained as shown in Fig.~\ref{fig:MLPNN_RBA}using the generic pre-generated training data as shown in TABLE~\ref{tab:rba_train_data} which was derived by following the flow diagram shown in Fig.~\ref{fig:RBA_flow} to get the FAs. LDM or MLP can be selected based on the computation efficiency. Furthermore, for RBA, a single LDM or MLP is tasked with identifying the connectivity of two consecutive bus-sections (nodes). Since the approach identifying the FA of the substation, all single breaker operations are considered similar to the the FA shown in Fig.~\ref{fig:func_vs_struct}. (f). Each LDM or MLP takes in the two consecutive bus section voltage phasor measurements and compare them and converts into logical values in pre-processing. The binary outputs of the pre-processing is used as the inputs to the LDM or MLP training data generation. The output is the connectivity in between the respective two bus-sections. All node connectivity statuses combined together to get the FA of the RBA substation as shown in Fig.~\ref{fig:MLPNN_RBA}. Equation~(\ref{eq:MLPNN_RBA}) shows the MLP that derive the connection between X and Y bus-sections (nodes) of RBA.

\begin{table}[!h]
\renewcommand{\arraystretch}{1}\
\caption{Training Dataset of MLP for X and Y node connectivity of RBA substation.}
\label{tab:rba_train_data}
\centering
\begin{tabular}{|ccc|c|}
\hline
\multicolumn{3}{|c|}{Inputs}                         & Output                         \\ \hline
\multicolumn{1}{|c|}{C} & \multicolumn{1}{c|}{A} & B & Node X and Y Connection Status \\ \hline
\multicolumn{1}{|c|}{1} & \multicolumn{1}{c|}{1} & 1 & 1                              \\ \hline
\multicolumn{1}{|c|}{1} & \multicolumn{1}{c|}{0} & 0 & 0                              \\ \hline
\multicolumn{1}{|c|}{0} & \multicolumn{1}{c|}{1} & 1 & 0                              \\ \hline
\multicolumn{1}{|c|}{0} & \multicolumn{1}{c|}{1} & 0 & 0                              \\ \hline
\multicolumn{1}{|c|}{0} & \multicolumn{1}{c|}{0} & 1 & 0                              \\ \hline
\multicolumn{1}{|c|}{0} & \multicolumn{1}{c|}{0} & 0 & 0                              \\ \hline
\end{tabular}
\end{table}

\begin{equation}
\label{eq:MLPNN_RBA}
Y_{X,Y} = f(A, B, C, B_1, [W_1], B_2, [W_2]) 
\end{equation}

\subsection{Single Bus Arrangement (SBA)}

\begin{figure}[h!]
\centering
\includegraphics[width=3.49in]{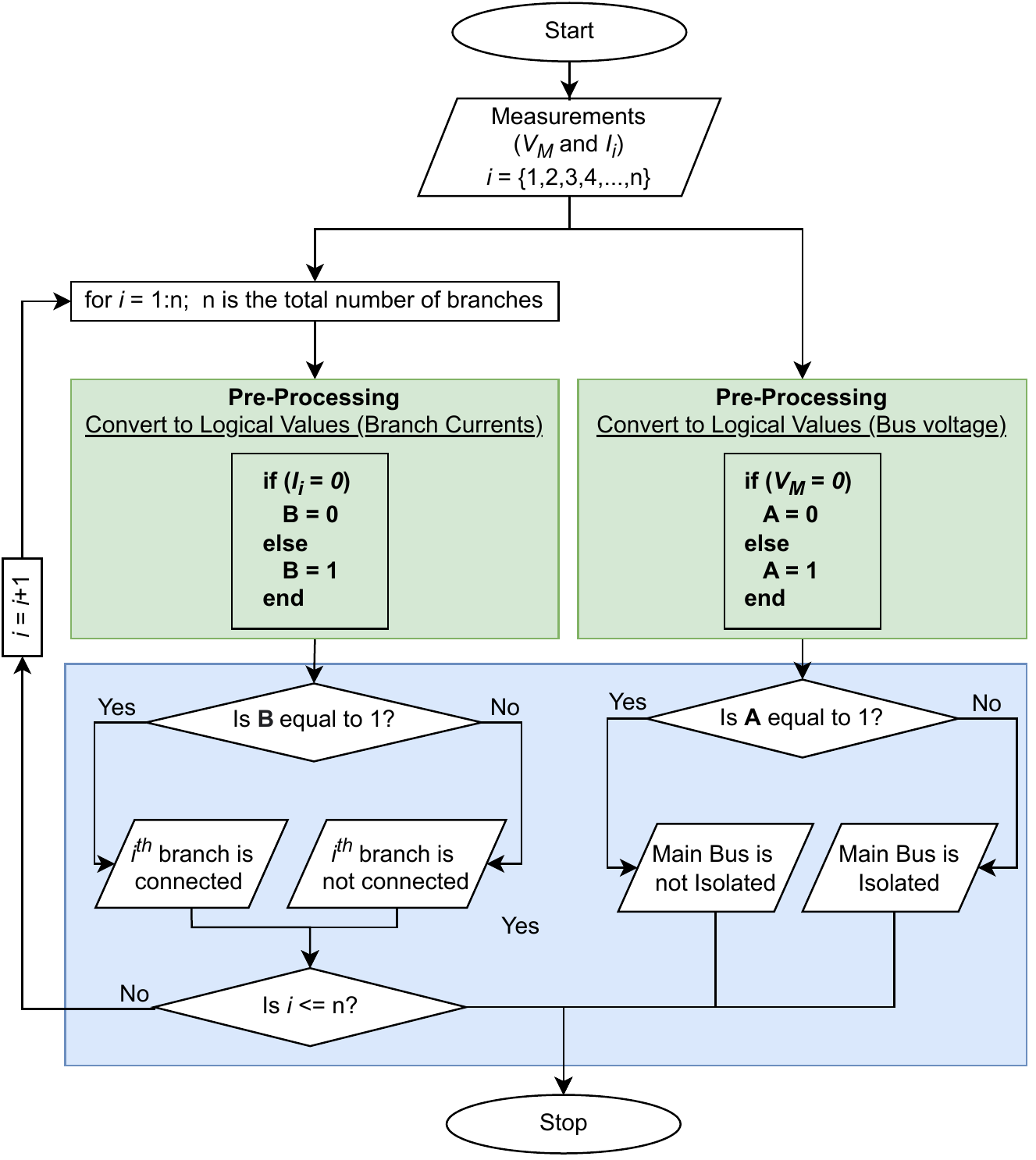}
\caption{A flow diagram for substation configuration identification of functional arrangement of SBA type substation.}
\label{fig:SBA_flow}
\end{figure}

SBA substation with six branches considered as the test case, which is shown in Fig.~\ref{fig:substations} (c). Branch current phasor measurements (circled in purple) and bus voltage phasor measurements (circled in orange) are in place as shown in Fig.~\ref{fig:substations} (c). All FA configurations of the SBA can be derived considering circuit nodal analysis using branch currents and the bus voltage. In the SBA substation, each branch current availability is considered to identify the branch connectivity to the bus which is connected through the breaker. It is important to defined the branch current availability rule based on the system expert knowledge about the current measurements for a no-load line vs disconnected branch. Combining the branch connection status with the bus voltage to identify the single bus isolation to confirm the substation is in service or not will be used to derive the Fa for the SBA. The SBA SCI method identify FAs as shown in Fig.~\ref{fig:SBA_flow} flow diagram. 

\begin{figure}[h!]
\centering
\includegraphics[width=3.49in]{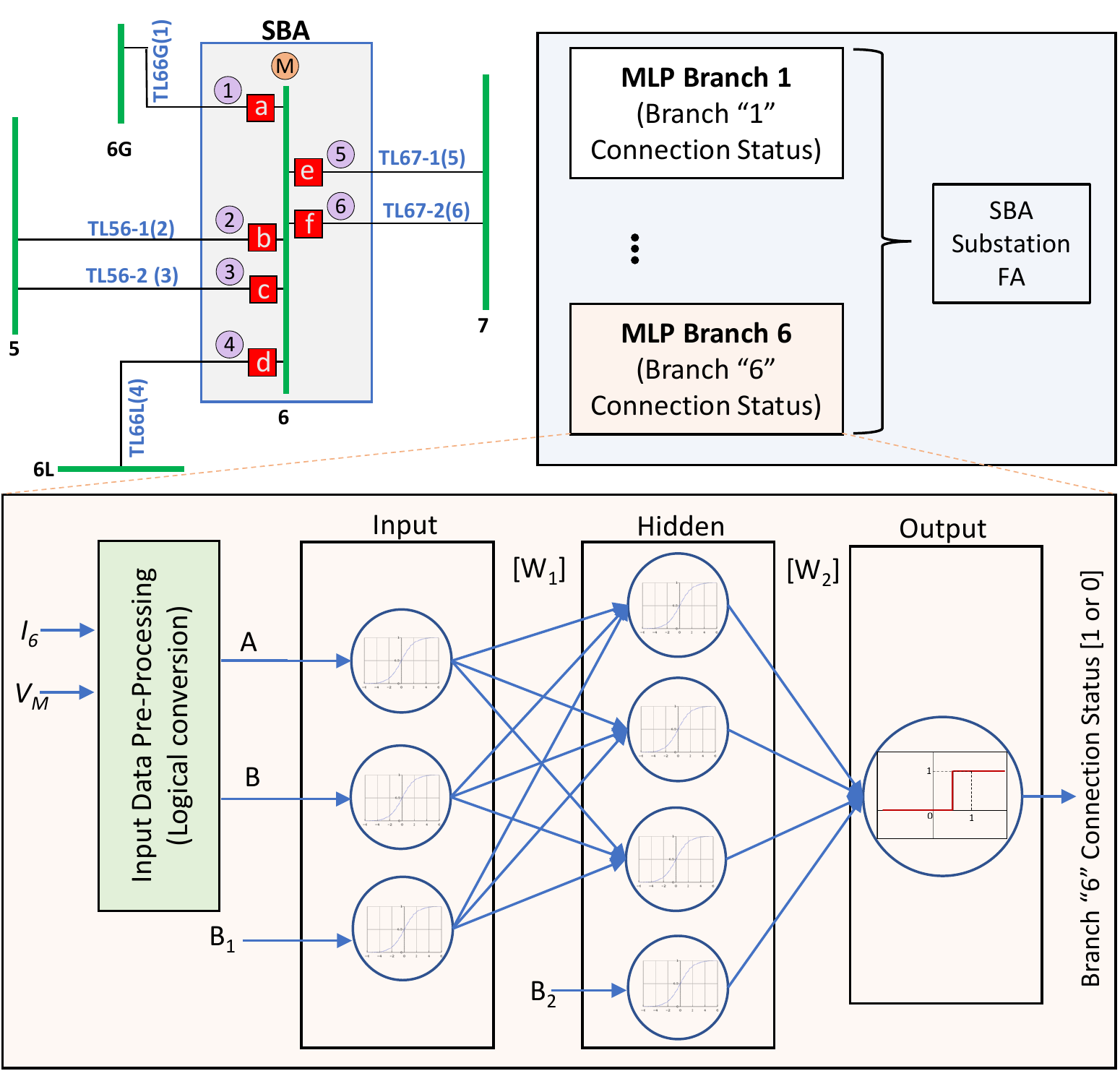}
\caption{The neural network structure of the MLP based substation configuration identification developed for six branch SBA substation shown in Fig.~\ref{fig:substations} (c).}
\label{fig:MLPNN_SBA}
\end{figure}

For SBA substation arrangement, an LDM can be setup by following the procedure shown in the Fig.~\ref{fig:SBA_flow} or a MLPs can be trained as shown in Fig.~\ref{fig:MLPNN_SBA} using the generic pre-generated training data as shown in TABLE~\ref{tab:sba_train_data} by following the flow diagram shown in Fig.~\ref{fig:SBA_flow} to identify FAs. Furthermore for SBA, each branch is considered by a single LDM or MLP, which takes bus voltage phasor and branch current phasor as input to identify the branch status and eventually derive the FAs as explained by the Fig.~\ref{fig:MLPNN_SBA}. MLP module use to identify the connectivity of $i^{th}$ branch of SBA can be summarized into~(\ref{eq:MLPNN_SBA}).

\begin{table}[!h]
\renewcommand{\arraystretch}{1}\
\caption{Training dataset of MLP for branch $i$ connectivity of SBA substation.}
\label{tab:sba_train_data}
\centering
\begin{tabular}{|cc|c|}
\hline
\multicolumn{2}{|c|}{Inputs} & Output                     \\ \hline
\multicolumn{1}{|c|}{B}  & A & Branch $i$ Connection Status \\ \hline
\multicolumn{1}{|c|}{1}  & 1 & 1                          \\ \hline
\multicolumn{1}{|c|}{0}  & 1 & 0                          \\ \hline
\multicolumn{1}{|c|}{0}  & 0 & 0                          \\ \hline
\end{tabular}
\end{table}

\begin{equation}
\label{eq:MLPNN_SBA}
Y_{i} = f(A, B, B_1, [W_1], B_2, [W_2]) 
\end{equation}


\section{Results \& Discussion}

\begin{figure*}[!ht]
    \centering
    \captionsetup{justification=centering,margin=2cm}
    \includegraphics[width=6.6in]{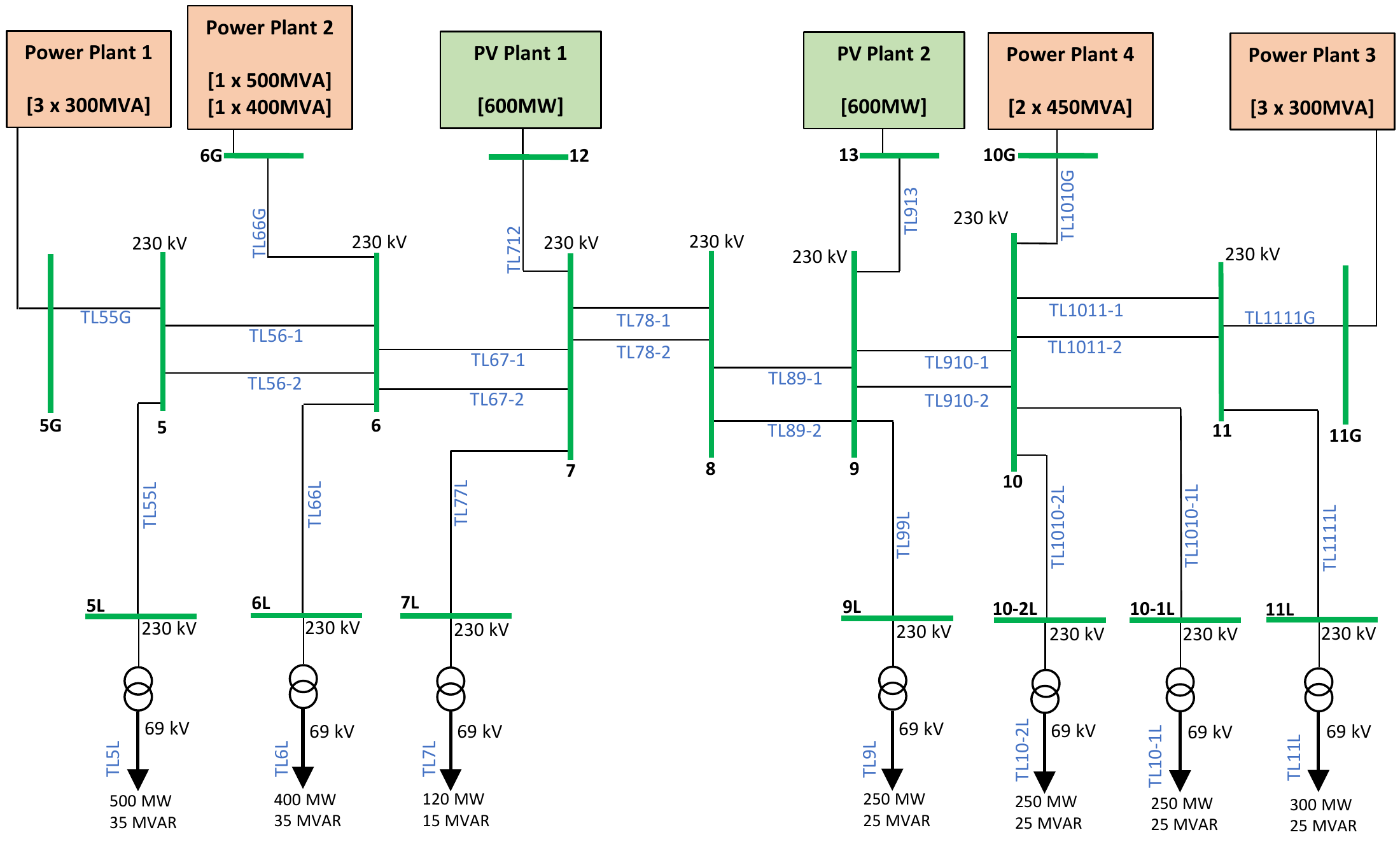}
    \caption{Test Case: Modified Two-Area Four-Machine Power System with renewable energy sources.}
    \label{fig:2a4m}
\end{figure*}

A modified two-area four-machine power system model is shown in Fig.~\ref{fig:2a4m}. The model consists of four conventional power plants at substation 5G, 6G, 10G and 11G and two RES based grid connected solar power plants at substation 12 and 13 instead of single generators in the original model~\cite{kundur}. The loads are distributed through load distributions substations 5L, 6L, 7L, 9L, 10L and 11L. Substation 5L is configured as MTBA, substation 6 configured as SBA and substation 5 is configured as RBA as shown in Fig.~\ref{fig:substations}. The simulation of the power system is carried out on the real-time digital simulator (RTDS). All measurements are obtained using PMUs.

MTBA, RBA and SBA substation arrangements are simulated on the RTDS simulator. The simulation PMU measurements is used to test the LDM and NN based SCIs proposed in section III. The SCI methods are developed in MATLAB. All possible substation FAs derived from the SCIs are cross-checked with the actual substation FAs in the test cases including unstable power system operation caused by some substation configurations, due to exceeding the (n-1) contingency level. Under all these operating conditions, substation configurations are checked. The SCI accuracy is not affected by the state of the power system. The TABLE~\ref{tab:accuracy} shows the SCI accuracy for the MTBA, RBA and SBA substation types. Computation time for LDM and NN for all three substation types estimated for single PMU frame processing is shown in TABLE~\ref{tab:timing}. The timing values are in the TABLE~\ref{tab:timing} are based on the same computational platform (The testing platform is a consist of Intel Xeon(R) Gold 3.3GHz with 63.7GB RAM). The computational timing is not a stable metric to estimate the computational efficiency, since it is based on the computational platform. Thus, a time ratio index is introduced. The time ratio index is the ratio between the LDM computation time and the NN computation time, which normalize the dependency on the computational platform. Based on the time ratio index, it is identified that for MTBA substation arrangement efficiency is close in both LDM and NN and for RBA and SBA substation types NN is efficient.

\begin{table}[!h]
\renewcommand{\arraystretch}{1}\
\caption{Substation Configuration Identification Accuracy of the LDM and NN approaches for the test cases MTBA. RBA and SBA substations shown in Fig.~\ref{fig:substations}.}
\label{tab:accuracy}
\centering
\begin{tabular}{|c|c|c|c|c|c|}
\hline
\begin{tabular}[c]{@{}c@{}}Substation\\ Type\end{tabular} & \begin{tabular}[c]{@{}c@{}}Substation\\ ID\end{tabular} & \begin{tabular}[c]{@{}c@{}}Number of \\ Breaker\end{tabular} & CAs & FAs & SCI Accuracy \\ \hline
MTBA                                                      & 5L                                                      & 3                                                            & 8   & 2   & 100\%        \\ \hline
RBA                                                       & 5                                                       & 4                                                            & 16  & 11  & 100\%        \\ \hline
SBA                                                       & 6                                                       & 6                                                            & 64  & 64  & 100\%        \\ \hline
\end{tabular}
\end{table}

\begin{table}[!h]
\renewcommand{\arraystretch}{1}\
\caption{Computation performance of the LDM and NN approaches for the test cases MTBA, RBA and SBA substations shown in Fig.~\ref{fig:substations}}
\label{tab:timing}
\centering
\begin{tabular}{|c|c|cc|c|}
\hline
\multirow{2}{*}{\begin{tabular}[c]{@{}c@{}}Substation \\ Type\end{tabular}} & \multirow{2}{*}{\begin{tabular}[c]{@{}c@{}}Substation \\ ID\end{tabular}} & \multicolumn{2}{c|}{Computation Time (us)} & \multirow{2}{*}{\begin{tabular}[c]{@{}c@{}}Time Ration Index\\ (t(LDM)/t(NN)\end{tabular}} \\ \cline{3-4}
                                                                            &                                                                           & \multicolumn{1}{c|}{LDM}        & NN    &                                                                                               \\ \hline
MTBA                                                                        & 5L                                                                        & \multicolumn{1}{c|}{27.98}       & 26.73    & 1.04                                                                                          \\ \hline
RBA                                                                         & 5                                                                         & \multicolumn{1}{c|}{77.35}      & 51.56    & 1.50                                                                                          \\ \hline
SBA                                                                         & 6                                                                         & \multicolumn{1}{c|}{130.45}     & 84.67    & 1.54                                                                                          \\ \hline
\end{tabular}
\end{table}

\begin{figure}[h!]
\centering
\includegraphics[width=3.3in]{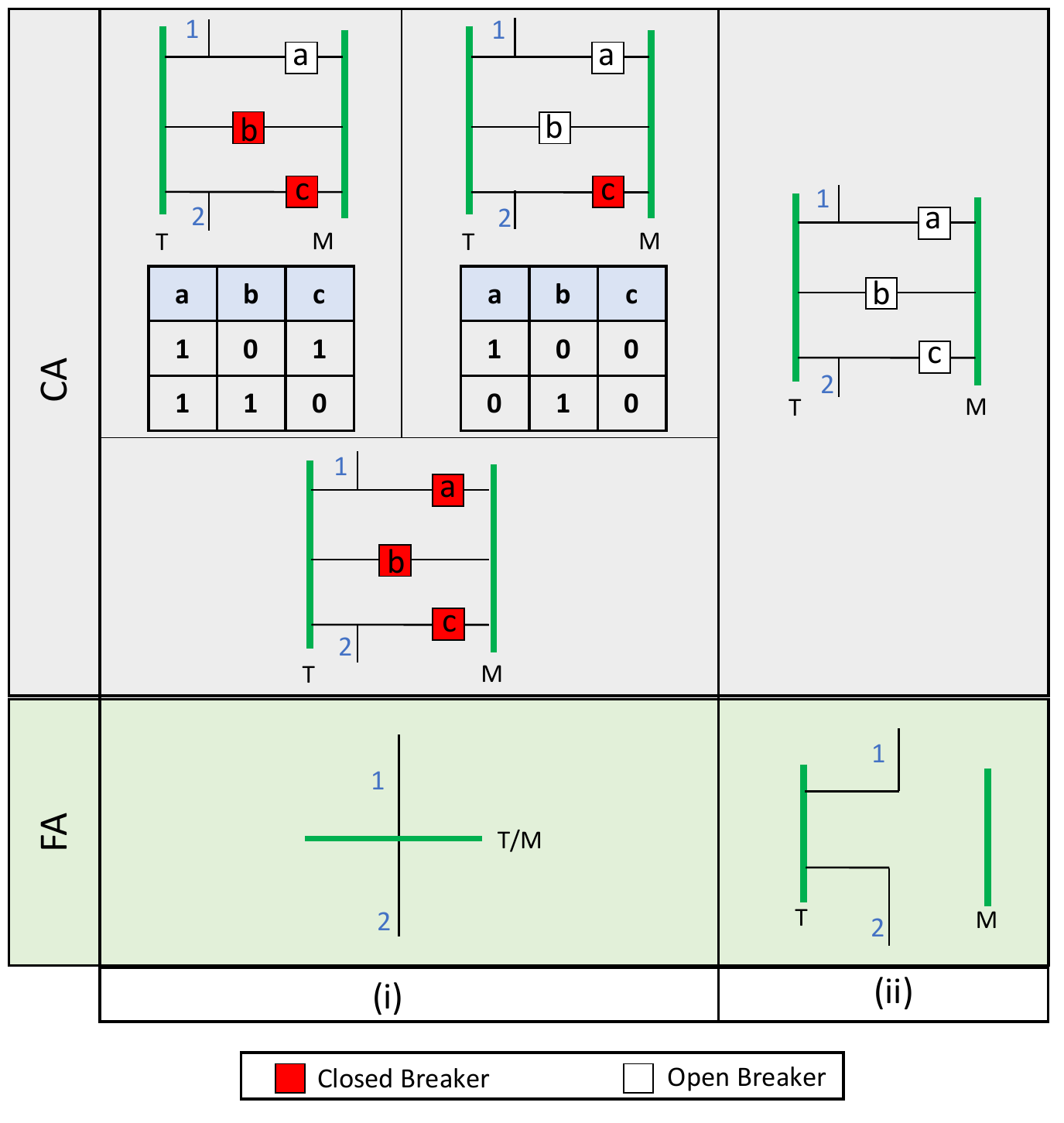}
\caption{Identified functional arrangements using proposed substation configuration identification approaches for the MTBA substation shown in Fig.~\ref{fig:substations} (a).}
\label{fig:result_1}
\end{figure}

The FAs for MTBA are shown in Fig.~\ref{fig:result_1}. There are two FAs. (i) case connects both Main and Transfer bus together and the (ii) case separate main and transfer bus and connect all the branches to the transfer bus, which isolate the main bus from the system.

\begin{figure}[h!]
\centering
\includegraphics[width=3.3in]{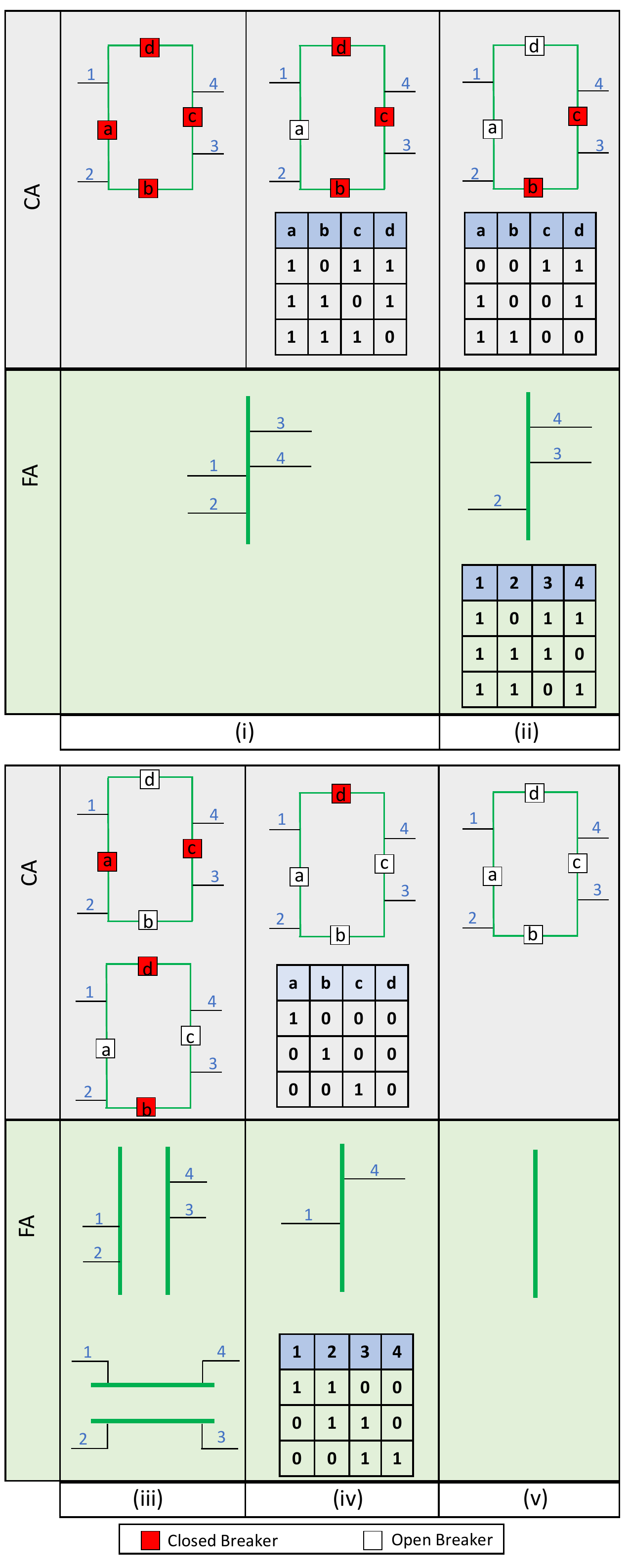}
\caption{Identified functional arrangements using proposed substation configuration identification approaches for the RBA substation shown in Fig.~\ref{fig:substations} (b).}
\label{fig:result_2}
\end{figure}

The FAs for RBA substation are shown in Fig.~\ref{fig:result_2}. There are total of twelve FAs possible for a four-node RBA substation. Which can be described under five categories as shown in Fig.~\ref{fig:result_2}. (i) category shows all branches are connected together through the RBA substation. This scenario can be occurred under five different CAs as shown, where all four breakers are closed or in any single breaker opened cases, which result in the same FA. (ii) category shows a single branch is disconnected from the substation. There are four possible CAs and respective four FAs in four-node RBA substation. Any two consecutive breakers can be opened to disconnect a single branch from the RBA substation. This characteristic of the RBA substation improve the substation reliability, where single breaker malfunction won't affect the power system operation. (iii) category shows system separation conducted by the RBA substation. This characteristics can be used for islanding part of the power system or limit cascading power system failures. In the four-node RBA, there are two possible system separation CAs and respective two FAs. These two separation can be done either opening only breakers "a" and "c" or breakers "b" and "d". (iv) category shows two consecutive branches were disconnected from the system in four-node RBA substation by opening three consecutive breakers. There are four possible CAs and respective four FAs. (v) category shows the four-node RBA substation is isolated from the power system by disconnecting all branches from the substation. This scenario can be occurred under major maintenance or a post blackout scenario. There is a single CA resulted under this category and the a respective single FA.

\begin{figure}[h!]
\centering
\includegraphics[width=3.3in]{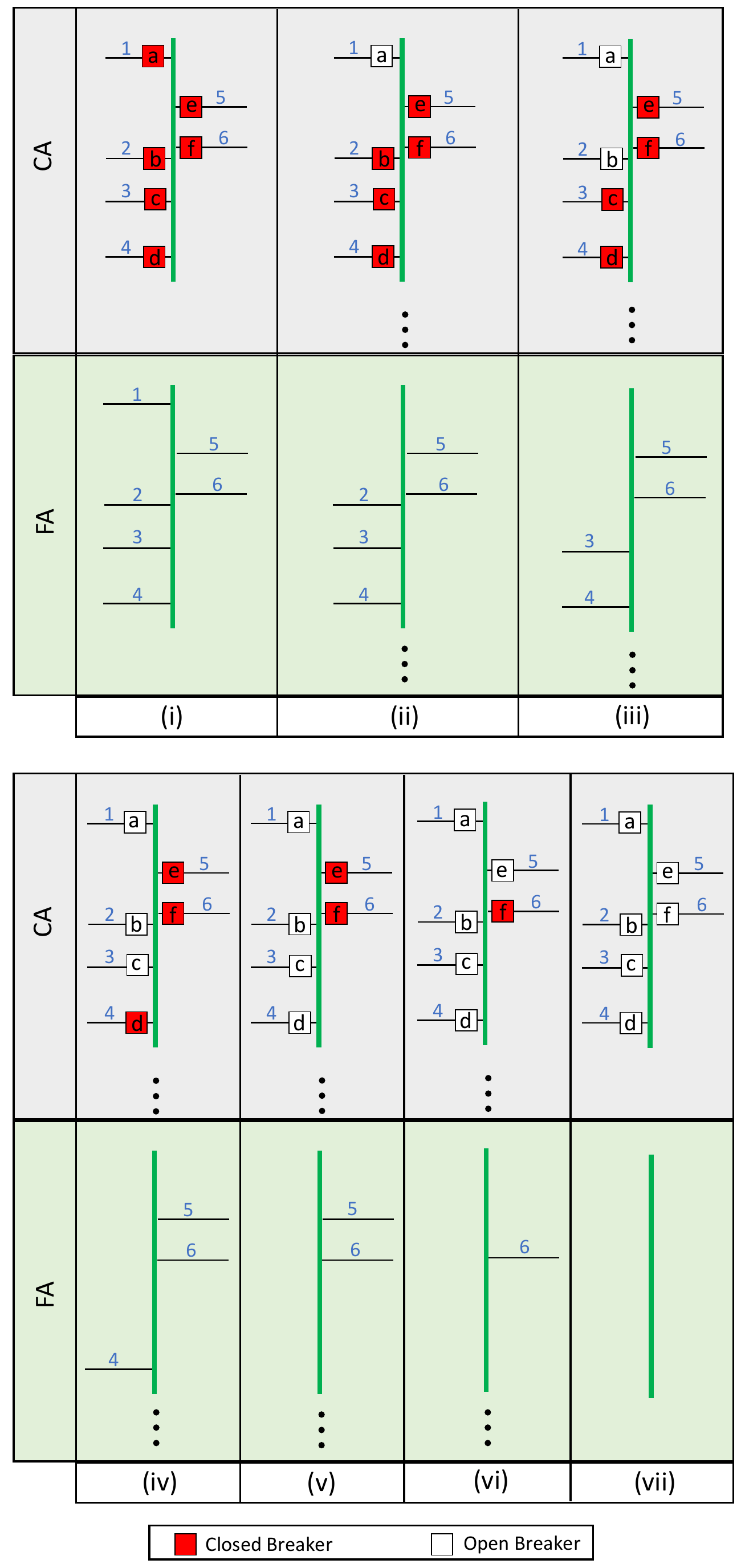}
\caption{Identified functional arrangements using proposed substation configuration identification approaches for the SBA substation shown in Fig.~\ref{fig:substations} (c).}
\label{fig:result_3}
\end{figure}

The FAs for SBA substation are shown in Fig.~\ref{fig:result_3}. Under SBA, all branches are connected through a common node. SBA arrangement is used in non-priority substation due to lack of reliability. There are 64 different CAs and respective 64 FAs in the six-branch SBA test case. As mentioned in the Section II - subsection C, SBA CAs are identical to FAs. These 64 FAs can be considered under seven categories as shown in Fig.~\ref{fig:result_3}. (i) category shows all the branches connected to the substation single bus. There is one possible FA. (ii) category shows single breaker is opened, which result in single branch disconnected from the substation. There are six possible FAs. (iii) category shows any two breakers are opened simultaneously, which result in two branches disconnected from the substation. There are 15 possible FAs. (iv) category shows any three breakers are opened simultaneously, which result in three branches disconnected from the substation. There are 20 possible FAs. (v) category shows any four breakers are opened simultaneously, which result in four branches disconnected from the substation. There are 15 possible FAs. (vi) category shows any five breakers are opened, which result in five branches disconnected from the substation. There are six possible FAs. (vii) category shows all six breakers are opened, which result in substation isolation from the power system. There is one possible FA.

\section{Conclusion}

Artificial intelligence based approaches for power system substation configuration identification for main and transfer bus arrangement, ring bus arrangement, and single bus arrangement has been presented in this paper. AI based approaches for substation configuration identifications were developed offline with the knowledge of the physical structure of the substation arrangement type and number of branches or bus-sections (nodes). The AI based substation configuration identifications were implemented and tested using data from phasor measurement units collected on a real-time simulation of a modified two-are four-machine power system model with two grid connected solar PV plants. The substation functional arrangements were identified accurately and fast for different power system operating states. Ongoing studies include extension of AI approaches for more complex substation arrangements and integrating the substation configuration identification into the transmission network topology processor for energy management system applications.





\bibliographystyle{IEEEtran}
%
\bibliography{mybib}


\end{document}